\documentclass{ws-procs9x6}

\usepackage{bm,epsfig}

\begin{document}

\title{Flavor Structure of Pentaquark Baryons in Quark Model}

\author{Yongseok Oh}

\address{Department of Physics and Astronomy, University of Georgia \\
Athens, Georgia 30602, USA\\
E-mail: yoh@physast.uga.edu}

\author{Hungchong Kim}

\address{Institute of Physics and Applied Physics, Yonsei University \\
Seoul 120-749, Korea\\
E-mail: hung@phya.yonsei.ac.kr}

\maketitle

\abstracts{
The flavor SU(3) group structure of pentaquark baryons which form
$\bm{1}$, $\bm{8}$, $\bm{10}$, $\overline{\bm{10}}$, $\bm{27}$, and
$\bm{35}$ multiplets is investigated in quark model.
The flavor wave functions of all the pentaquark baryons are
constructed in SU(3) quark model and their Yukawa interactions with
meson octet are obtained in general and in the special case of the
octet-antidecuplet ideal mixing with the OZI rule.
The mass sum rules of pentaquark baryons are also discussed.
}

\section{Introduction}

Great interests in exotic baryons in hadron physics have been initiated
by the discovery of $\Theta^+(1540)$ state by the LEPS Collaboration and
subsequent experiments~\cite{LEPS03,theta:pos}.
The observation of $\Xi^{--}(1862)$ by NA49 Collaboration~\cite{NA49-03}
may suggest that $\Xi(1862)$ forms pentaquark antidecuplet with
$\Theta^+(1540)$ as anticipated by the soliton model study~\cite{DPP97}.
Later, the H1 Collaboration reported the existence of anti-charmed
pentaquark state~\cite{H1-04}, which revives the interests in the heavy
pentaquark system~\cite{Lip87-GSR87,heavy,KLO04}.
However, the existence of pentaquark baryons are not fully 
confirmed by experiments yet as some high energy experiments report null
results for those states~\cite{theta:null,xi:null}.
A summary for the experimental situation and perspectives can be found,
e.g., in Refs.~\refcite{CLAS04a}.
Theoretically, many ideas have been suggested and developed to study the
exotic pentaquark states in various approaches and
models~\cite{JM03,Jaffe04}, but more detailed studies are required to
understand the properties and formation of pentaquark states.

As the pentaquark baryons may be produced in photon-hadron or
hadron-hadron reactions, it is important to understand their production
mechanisms and decay channels in order to confirm the existence of the
pentaquark states and to study their properties.
The present studies on the production reactions are limited by the lack
of experimental and phenomenological inputs on some
couplings~\cite{LK03,OKL03,DKST03-Roberts04,NT03,NHK04}.
In particular, those studies could not include the contributions from the
intermediate pentaquark states in production mechanisms.
Therefore, it is strongly desired to understand the interactions of
pentaquark baryons with other hadrons.

On the other hand, many theoretical speculations suggest that the physical
pentaquark states would be mixtures of various
multiplets~\cite{JW03,CD04,EKP04}.
Thus it is necessary to construct the wavefunctions of pentaquark
baryons in terms of quark and antiquark for understanding the structure of
pentaquark states.
In this talk, we discuss a way to construct the wavefunctions of all
the pentaquark baryons in quark model and obtain their SU(3) symmetric
interactions with other baryons.
Then several mass relations among the pentaquark baryons are discussed.
In addition, we explore the couplings in the special case of the
antidecuplet-octet ideal mixing with the OZI rule.
The topics presented here are discussed in more detail in
Refs.~\refcite{OKL03b,LKO04,OK04}.

\section{Wavefunctions and interactions of pentaquark baryons}

We start with the representations for quark and antiquark.
We denote a quark by $q_i$ and an antiquark by $q^i$ with $i=1,2,3$, so
that $q_1$, $q_2$, and $q_3$ are $u$, $d$, and $s$ quark, respectively.
The inner products of the quark and antiquark operators are normalized as
\begin{equation}
(q_i, q_j) = \delta_{ij}, \qquad (q^i, q^j) = \delta^{ij}, \qquad
(q_i, q^j) = 0.
\end{equation}

Then the diquark state is decomposed as
\begin{equation}
q_j q_k = \frac{1}{\sqrt2} S_{jk} + \frac{1}{2\sqrt2} \epsilon_{ijk}
T^i,
\end{equation}
where
\begin{equation}
\left( \begin{array}{c} S_{jk} \\ A_{jk} \end{array} \right) =
\frac{1}{\sqrt2} \left( q_j q_k \pm q_k q_j \right),
\end{equation}
and
\begin{equation}
T^i = \epsilon^{ijk} A_{jk},
\end{equation}
so that $S_{jk}$ and $T^i$ represent $\bm{6}$ and $\overline{\bm{3}}$,
respectively. This shows that $\bm{3} \otimes \bm{3} = \bm{6} \oplus
\overline{\bm{3}}$.

The product of two diquarks can be written as
\begin{eqnarray}
(q_j^{} q_k^{}) (q_l^{} q_m^{}) &=& \frac{1}{2\sqrt6} T_{jklm}
+ \frac{1}{4\sqrt2} \left( \epsilon_{akl} \delta^b_m \delta^c_j +
\epsilon_{ajm} \delta^b_l \delta^c_k \right) S^a_{bc}
\nonumber \\ && \mbox{}
+ \frac{1}{\sqrt6} \left( \epsilon_{ajl} \epsilon_{bkm} +
\epsilon_{akl}\epsilon_{bjm} \right) T^{ab}
\nonumber \\ && \mbox{}
+ \frac14 \epsilon_{ijk} \left\{ T^i_{lm} + \frac{1}{\sqrt2} \left(
\delta^i_l \delta^a_m + \delta^i_m \delta^a_l \right) Q_a \right\}
\nonumber \\ && \mbox{}
+ \frac14 \epsilon_{ijk} \left\{ \widetilde{T}^i_{lm} + \frac{1}{\sqrt2}
\left(\delta^i_l \delta^a_m + \delta^i_m \delta^a_l \right)
\widetilde{Q}_a \right\}
\nonumber \\ && \mbox{}
+ \frac{1}{4\sqrt2} \epsilon_{ijk} \epsilon_{nlm} \left( 2 S^{in} +
\epsilon^{ina} T_a \right).
\label{eq:twodq}
\end{eqnarray}
Here, $T_a$, $Q_a$, and $\widetilde{Q}_a$ are $(1,0)$ type and
represent $\bm{3}$.
$T^{ij}$ and $S^{ij}$ are $(0,2)$ type of $\overline{\bm{6}}$.
Also $T^{i}_{jk}$, $\widetilde{T}^i_{jk}$, and $S^{i}_{jk}$ are $(2,1)$
type of $\bm{15}$ and $T_{ijkl}$ are $(4,0)$ type of $\bm{15}$.
Their explicit forms can be found in Ref.~\refcite{OK04}.

Then it is straightforward to obtain the wavefunctions of pentaquark
baryons. Since the product of two diquarks can form $\bm{3}$,
$\overline{\bm{6}}$, and $\bm{15}$, the pentaquark states can have
$\bm{1}$, $\bm{8}$, $\bm{10}$, $\overline{\bm{10}}$, $\bm{27}$, and
$\bm{35}$, while the normal three-quark baryons have $\bm{8}$ and
$\bm{10}$. So the pentaquark states have much richer spectrum than the
normal baryons. Their flavor wavefunctions are obtained by direct product of
$(q_j^{} q_k^{}) (q_l^{} q_m^{})$ and an antiquark $\bar{q}^a$.
For example, the $\bm{35}$-plet tensors $T^a_{ijkl}$ and the
$\bm{27}$-plet tensors $T^{ij}_{kl}$ can be constructed as
\begin{eqnarray}
T^a_{ijkl} &=& T_{ijkl} \bar{q}^a - \frac{1}{6} \left( \delta^a_i
T_{jklm} \bar{q}^m + \delta^a_j T_{iklm} \bar{q}^m +
\delta^a_k T_{ijlm} \bar{q}^m + \delta^a_l T_{ijkm} \bar{q}^m
\right),
\nonumber \\
T^{ij}_{kl} &=& c_1^{} \left\{
\frac{1}{2\sqrt{2}} \left( T^i_{kl} \bar{q}^j + T^j_{kl}
\bar{q}^i \right)
\right. \nonumber \\ && \mbox{} \left. \qquad
- \frac{1}{10\sqrt{30}} \left( \delta^i_l T^j_{km}
\overline{q}^m + \delta^j_l T^i_{km} \overline{q}^m +
\delta^i_k T^j_{lm} \overline{q}^m
+ \delta^j_k T^i_{lm} \overline{q}^m \right) \right\}
\nonumber \\ && \mbox{}
+ c_2^{} \left\{
\frac{1}{2\sqrt{2}} \left( \widetilde{T}^i_{kl} \bar{q}^j + \widetilde{T}^j_{kl}
\bar{q}^i \right)
\right. \nonumber \\ && \mbox{} \left. \qquad
- \frac{1}{10\sqrt{30}} \left( \delta^i_l \widetilde{T}^j_{km}
\overline{q}^m + \delta^j_l \widetilde{T}^i_{km} \overline{q}^m
+ \delta^i_k \widetilde{T}^j_{lm} \overline{q}^m
+ \delta^j_k \widetilde{T}^i_{lm} \overline{q}^m \right) \right\}
\nonumber \\ && \mbox{}
+ c_3^{} \left\{
\frac{1}{2\sqrt{2}} \left( S^i_{kl} \bar{q}^j + S^j_{kl}
\bar{q}^i \right)
\right. \nonumber \\ && \mbox{} \left. \qquad
- \frac{1}{10\sqrt{30}} \left( \delta^i_l S^j_{km}
\overline{q}^m + \delta^j_l S^i_{km} \overline{q}^m +
\delta^i_k S^j_{lm} \overline{q}^m
+ \delta^j_k S^i_{lm} \overline{q}^m \right) \right\}.
\nonumber \\
\end{eqnarray}
The pentaquark octet $P^i_j$ and antidecuplet $T^{ijk}$ read
\begin{eqnarray}
P^i_j &=& \frac{c_1^{}}{\sqrt2} \left( T_j \overline{q}^i - \frac13
\delta^i_j T_m \overline{q}^m \right)
+ \frac{c_2^{}}{\sqrt2} \left( Q_j \overline{q}^i - \frac13
\delta^i_j Q_m \overline{q}^m \right)
\nonumber \\ && \mbox{}
+ \frac{c_3^{}}{\sqrt2} \left( \widetilde{Q}_j \overline{q}^i - \frac13
\delta^i_j \widetilde{Q}_m \overline{q}^m \right)
+ \frac{c_4^{}}{\sqrt3} \epsilon_{jab} S^{ia} \overline{q}^b
+ \frac{c_5^{}}{\sqrt3} \epsilon_{jab} T^{ia} \overline{q}^b
\nonumber \\ && \mbox{}
+ \frac{c_6^{}}{\sqrt{15}} T^i_{jk} \overline{q}^k
+ \frac{c_7^{}}{\sqrt{15}} \widetilde{T}^i_{jk} \overline{q}^k
+ \frac{c_8^{}}{\sqrt{15}} S^i_{jk} \overline{q}^k,
\nonumber \\
T^{ijk} &=& \frac{c_1^{}}{\sqrt3} \left( S^{ij} \overline{q}^k
+ S^{jk} \overline{q}^i + S^{ki} \overline{q}^j \right)
+ \frac{c_2^{}}{\sqrt3} \left( T^{ij} \overline{q}^k
+ T^{jk} \overline{q}^i + T^{ki} \overline{q}^j \right),
\end{eqnarray}
and the pentaquark decuplet $D_{ijk}$ and the singlet $S$ are
\begin{eqnarray}
D_{ijk} &=& \frac{c_1^{}}{\sqrt6} T_{ijkl} \overline{q}^l
+ \frac{c_2^{}}{\sqrt{24}} \left ( \epsilon_{iab} T^a_{jk} \overline{q}^b
+\epsilon_{jab} T^a_{ki} \overline{q}^b+\epsilon_{kab} T^a_{ij} \overline{q}^b
\right )
\nonumber \\ && \mbox{}
+ \frac{c_3^{}}{\sqrt{24}} \left ( \epsilon_{iab} \widetilde{T}^a_{jk}
\overline{q}^b+\epsilon_{jab} \widetilde{T}^a_{ki}
\overline{q}^b+\epsilon_{kab} \widetilde{T}^a_{ij}
\overline{q}^b \right )
\nonumber \\ && \mbox{}
+ \frac{c_4^{}}{\sqrt{24}} \left (
\epsilon_{iab} S^a_{jk} \overline{q}^b
+\epsilon_{jab} S^a_{ki} \overline{q}^b
+\epsilon_{kab} S^a_{ij} \overline{q}^b
\right ) ,
\nonumber \\
S &=& - \frac{c_1^{}}{\sqrt6} T_m \overline{q}^m
 - \frac{c_2^{}}{\sqrt6} Q_m \overline{q}^m
 - \frac{c_3^{}}{\sqrt6} \widetilde{Q}_m \overline{q}^m.
\end{eqnarray}

Therefore, by constructing all possible pentaquark tensors we can verify 
\begin{equation}
\bm{3} \otimes \bm{3} \otimes \bm{3} \otimes \bm{3} \otimes
\overline{\bm{3}} = \bm{35} \oplus (3)\bm{27} \oplus
(2)\overline{\bm{10}} \oplus
(4)\bm{10} \oplus (8)\bm{8} \oplus (3)\bm{1},
\end{equation}
where the numbers in parentheses are the number of multiplicity.
The inner products of the multiplets are given in Ref.~\refcite{OK04}.
With those informations at hand, one can identify the tensor
representations with the baryon states of definite isospin and
hypercharge.
(See Ref.~\refcite{OK04} for details.)

The SU(3) symmetric Yukawa interactions of pentaquarks can be
constructed by fully contracting the
upper and lower indices of the three tensors representing two baryon
multiplets and the meson octet.
When the number of upper indices does not match that of lower indices,
the Levi-Civita tensors $\epsilon_{ijk}$ are introduced to
make the interactions fully contracted.
The SU(3) symmetric interactions constructed in this way give several
constraints or selection rules to the pentaquark interactions, which should
be useful to identify the pentaquark states.
Since
\begin{eqnarray}
\bm{8} \otimes \bm{8} &=& \bm{27} \oplus \bm{10} \oplus \overline{\bm{10}}
\oplus \bm{8}_1 \oplus \bm{8}_2 \oplus \bm{1}, \nonumber \\
\bm{10} \otimes \bm{8} &=& \bm{35} \oplus \bm{27} \oplus \bm{10} \oplus
\bm{8}, \nonumber \\
\overline{\bm{10}} \otimes \bm{8} &=& \overline{\bm{35}} \oplus \bm{27}
\oplus \overline{\bm{10}} \oplus \bm{8},
\nonumber \\
\bm{27} \otimes \bm{8} &=& \bm{64} \oplus \bm{35} \oplus
\overline{\bm{35}} \oplus \bm{27}_1 \oplus \bm{27}_2 \oplus \bm{10} \oplus
\overline{\bm{10}} \oplus \bm{8},
\nonumber \\
\bm{35} \otimes \bm{8} &=& \bm{81} \oplus \bm{64} \oplus \bm{35}_1 \oplus
\bm{35}_2 \oplus \bm{28} \oplus \bm{27} \oplus \bm{10},
\label{eq:int}
\end{eqnarray}
we find the followings.
First, the pentaquark singlet can couple to pentaquark octet only.
Second, the $\bm{27}$-$\bm{27}$ and $\bm{35}$-$\bm{35}$ interactions
have two types ($f$ and $d$ types) like $\bm{8}$-$\bm{8}$ interaction.
Third, the interactions including $\bm{10}$-$\overline{\bm{10}}$,
$\bm{35}$-$\bm{8}$, and $\bm{35}$-$\overline{\bm{10}}$ are not allowed
as they cannot form SU(3)-invariant interactions.
Thus, $\bm{35}$-plet couplings are limited to the interactions with
$\bm{27}$-plet and decuplet.

We refer to Refs.~\refcite{OKL03b,OK04} for the explicit relations for
pentaquark interactions.
The SU(3) symmetry breaking terms can be included in a standard
way~\cite{LKO04,PS04,GS04}.

\section{Mass sum rules}

Since all the particles belonging to an irreducible representation of
SU(3) are degenerate in the SU(3) symmetry limit, it is required to
include SU(3) symmetry breaking to obtain the mass splittings.
It is well-known that the Hamiltonian which breaks SU(3) symmetry but
still preserves the isospin symmetry and hypercharge is proportional to
the Gell-Mann matrix $\lambda_8$, from which we introduce the hypercharge
tensor as $\mathcal{Y} = \mbox{diag}( 1 , 1,-2)$.
Then the baryon masses can be obtained by constructing all possible
contractions among irreducible tensors and the hypercharge tensor.
As the mass formulas contain several parameters which take different
values depending on the multiplet in general, we can obtain only the
mass relations.
The Hamiltonian constructed in this way reads
\begin{eqnarray}
H_{\bf 8} &=& a \overline{P}^i_j P^j_i + b\overline{P}^i_j \mathcal{Y}^l_i
P^j_l + c \overline{P}^i_j \mathcal{Y}^j_l P^l_i,
\nonumber \\
H_{\bf 10} &=& a \overline{D}^{ijk} D_{ijk} + b \overline{D}^{ijk}
\mathcal{Y}^l_k D_{ijl},
\nonumber \\
H_{\bf \overline{10}} &=& a \overline{T}_{ijk} T^{ijk} +
b \overline{T}_{ijk} \mathcal{Y}_l^k T^{ijl},
\nonumber \\
H_{\bf 27} &=& a \overline{T}^{ij}_{kl} T^{kl}_{ij} +
b \overline{T}^{ij}_{kl} \mathcal{Y}^l_m T^{km}_{ij} +
c \overline{T}^{ij}_{kl} \mathcal{Y}^m_j T^{kl}_{im},
\nonumber \\
H_{\bf 35} &=& a \overline{T}^{jklm}_i T^{i}_{jklm} +
b \overline{T}^{jklm}_i \mathcal{Y}^n_j T^{i}_{nklm} +
c \overline{T}^{jklm}_i \mathcal{Y}^i_n T^{n}_{jklm},
\end{eqnarray}
where $a$, $b$, and $c$ are mass parameters.
Then, in addition to the well-known Gell-Mann--Okubo mass relation for
the baryon octet and the decuplet equal-spacing rule, we have some
interesting mass sum rules for antidecuplet, $\bm{27}$-plet, and
$\bm{35}$-plet.
In antidecuplet, we have the equal spacing rule~\cite{DPP97},
\begin{equation}
\Xi_{\overline{10},3/2} - \Sigma_{\overline{10}} =
\Sigma_{\overline{10}} - N_{\overline{10}} =
N_{\overline{10}} - \Theta.
\end{equation}
In the $\bm{27}$-plet, we find the analog of the Gell-Mann--Okubo
mass relation,
\begin{eqnarray}
2(N_{27}+\Xi_{27}) = 3\Lambda_{27} +\Sigma_{27}.
\end{eqnarray}
In addition, we find that some of the {\bf 27}-plet members, i.e.,
$\Theta_1$,
$\Delta_{27}$, $\Sigma_{27,2}$, $\Xi_{27,3/2}$, and $\Omega_{27,1}$,
satisfy two independent equal-spacing rules,
\begin{eqnarray}
\Omega_{27,1} - \Xi_{27,3/2} &=&
\Xi_{27,3/2} - \Sigma_{27,2},
\nonumber \\
\Sigma_{27,2} - \Delta_{27} &=&
\Delta_{27} - \Theta_1.
\end{eqnarray}
Note that they are the states with maximum isospin for a given
hypercharge and the equal-spacing rule holds independently for the upper
half of the $\bm{27}$-plet weight diagram and for the lower half of that
weight diagram~\cite{OK04}.

For the $\bm{35}$-plet baryons, we observe that there are two sets
of baryons which satisfy the equal-spacing rule separately~\cite{OK04},
namely,
\begin{eqnarray}
&&
\Omega_{35} - \Xi_{35} = \Xi_{35} - \Sigma_{35} =
\Sigma_{35} - \Delta_{35} = \Delta_{35} - \Theta_2,
\nonumber \\ &&
X - \Omega_{35,1} = \Omega_{35,1} - \Xi_{35,3/2} = \Xi_{35,3/2} -
\Sigma_{35,2} = \Sigma_{35,2} - \Delta_{5/2}.
\end{eqnarray}

\section{Ideal mixing of antidecuplet and octet with the OZI rule}

In the diquark-diquark-antiquark model for pentaquarks, Jaffe and Wilczek
advocated the ideal mixing of the antidecuplet with the octet~\cite{JW03}.
By referring the detailed discussion on the ideal mixing and the OZI rule to
Ref.~\refcite{LKO04}, here we discuss the consequence of the OZI rule in
the interactions of pentaquark octet and antidecuplet.
As can be seen from Eq.~(\ref{eq:int}), the pentaquark octet interaction
with normal baryon octet and meson octet has two couplings, $f$ and $d$.
A relation between the two couplings can be found by imposing the OZI
rule or the fall-apart mechanism~\cite{CD04}.
To see this, we go back to Eq.~(\ref{eq:twodq}) and note that the pentaquark
octet and antidecuplet come together from the $\overline{\bf 6}$ of two
diquarks and $\overline{\bf 3}$ of one antiquark, i.e.,
$\overline {\bf 6} \otimes \overline {\bf 3} = \overline {\bf 10}
\oplus {\bf 8}$.
This follows from
\begin{eqnarray}
S^{ij} \otimes \bar{q}^k = T^{ijk} \oplus S^{[ij,k]}.
\label{tensor1}
\end{eqnarray}
Obviously, the last part, being an octet representation, can be 
replaced by a two-index field $P^j_i$ such as
\begin{eqnarray}
S^{[ij,k]}=\epsilon^{ljk} P_l^i+\epsilon^{lik}P_l^j.
\label{tensor2}
\end{eqnarray}
In this scheme, the pentaquark antidecuplet and pentaquark
octet have the same universal coupling constant.
It is now clear to see that 
the index $k$ in Eq.~(\ref{tensor2}), the index for the antiquark, should
be contracted with the antiquark index of the meson field to represent
the fall-apart
mechanism or the OZI rule, as the usual baryon $B$ does not contain an
antiquark in the OZI limit.
Hence, the interaction should follow the form as
\begin{eqnarray}
\mathcal{L}_{\rm int}= g_{8}^{} \epsilon^{ilm} \overline{S}_{[ij,k]}
B^j_l M_m^k + \mbox{(H.c.)}.
\label{interaction2}
\end{eqnarray}
Substituting Eq.~(\ref{tensor2}) into Eq.~(\ref{interaction2}), one has
\begin{eqnarray}
\mathcal{L}_{\rm int} = 2 g_8^{} \overline{P}^m_i B^i_l M^l_m + g_8
\overline{P}^m_i M^i_l B^l_m+ \mbox{(H.c.)}.
\end{eqnarray}
Comparison with the standard expression for the octet baryon
interactions of $f$ and $d$ types leads to $f=1/2$ and $d=3/2$.
Therefore, one can find that the OZI rule makes a special choice on the
$f/d$ ratio as $f/d=1/3$~\cite{CD04,LKO04}.

\section{Summary}

We have obtained the flavor wavefunctions of all the pentaquark baryons
in quark model. Then the SU(3) symmetric interactions of the pentaquark
baryons as well as their mass sum rules are derived.
This will help to identify not only exotic baryons but also
crypto-exotic states.
At this stage, we notice that there are several recent reports about the
existence of crypto-exotic pentaquark states~\cite{oexotic}, whose
existence, however, should be clarified by further experiments~\cite{ZACC04}.

\section*{Acknowledgments}

We are grateful to Su Houng Lee for useful discussions.
This work was supported in part by Forschungszentrum-J{\"u}lich, contract
No. 41445282 (COSY-058) and the Brain Korea 21 project of Korean
Ministry of Education.

\end{document}